# An Intuitive Approach to Inertial Sensor Bias Estimation

Vasiliy M. Tereshkov

**Abstract:** A simple approach to gyro and accelerometer bias estimation is proposed. It does not involve Kalman filtering or similar formal techniques. Instead, it is based on physical intuition and exploits a duality between gimbaled and strapdown inertial systems. The estimation problem is decoupled into two separate stages. At the first stage, inertial system attitude errors are corrected by means of a feedback from an external aid. In the presence of uncompensated biases, the steady-state feedback rebalances those biases and can be used to estimate them. At the second stage, the desired bias estimates are expressed in a closed form in terms of the feedback signal. The estimator has only three tunable parameters and is easy to implement and use. The tests proved the feasibility of the proposed approach for the estimation of low-cost MEMS inertial sensor biases on a moving land vehicle.

---

## 1. INTRODUCTION

In most inertial measurement devices, e.g., attitude and heading reference systems (AHRS), the output performance suffers from gyro and accelerometer biases. Uncompensated portions of these biases result in unbounded accumulation of attitude errors. A classical approach to AHRS correction, dated back to the 1930s, is attitude error elimination by means of an external aid that provides vehicle acceleration data [1], [2]. In this scheme, the differences between the accelerometer measurements and the true accelerations are treated as gravity projections proportional to the AHRS stable platform tilt errors. The platform gyros are torqued until these errors vanish and the platform returns to the level plane.

This approach is also directly applicable to a strapdown AHRS if a notion of a "virtual platform" is introduced [3], [4]. The direction cosine matrix updated by the AHRS computer represents the orientation of the vehicle relative to a "virtual platform", and the angular motion of the vehicle is the difference of its absolute rotation and the motion of that "platform".

Even though the correction procedure prevents the accumulation of attitude errors, it cannot completely zero them, as inertial sensor biases remain uncompensated.

In modern studies, Kalman filtering [5] is usually preferred as a more general and powerful approach to aided inertial system design. The state vector to be estimated is composed of output errors and then, if necessary, is augmented by inertial sensor biases [6], [7]. Thus the phases of attitude correction and sensor error estimation are replaced by a single update procedure of the Kalman filter.

Though very popular, Kalman filtering has some inherent drawbacks. First, for any time-varying or nonlinear system (an AHRS being among them) it requires on-line computation of the error covariance matrix, which squares the number of necessary update operations at each time step. Second, the Jacobians of dynamics and measurement functions must be provided. Third, the optimality and even the convergence of the filter are not guaranteed for nonlinear systems [8]. Fourth, the formal nature of the Kalman filter makes it non-intuitive and difficult to tune [9].

As sensor biases are very slow varying quantities with respect to vehicle attitude parameters, a "separate-bias" estimation technique was proposed [10], [11] that decouples the filter into two stages, thus reducing the computational burden. The first stage provides the "bias-free" state estimates. The second stage estimates the biases in terms of the residuals in those "bias-free" estimates. The algorithm can be further simplified if the filter gains are calculated in a deterministic way [12], so that the covariance matrix and the Jacobians are no longer needed. The "separate-bias" framework is general enough but the question remains of how to select the appropriate dynamics model and filter gains to obtain convergent and easy to tune estimation for a given nonlinear system.

The dissatisfaction with existing filtering methods leads to the growing interest in estimation approaches which make use of some specific properties of inertial navigation problems [14], [15]. These approaches, sometimes called "symmetry-preserving filters", are essentially the implementations of the "virtual platform" technique treated from the group theoretical standpoint. "Symmetry-preserving filters" usually include gyro bias estimators. Accelerometer bias estimation, however, cannot be naturally incorporated into the filter structure.

In this paper the bias estimation problems for gyros and accelerometers are treated in a unified manner. The key idea is that attitude error correction equations of a strapdown AHRS can be considered as the first stage of a "separate-bias" estimator, and the residual "torques" applied to the "virtual platform" can be directly used as measurements for an extremely simple inertial sensor

---

Vasiliy M. Tereshkov is with Topcon Positioning Systems, LLC, 7/22 Derbenevskaya Embankment, Moscow, Russia (VTereshkov@topcon.com).

bias estimator.

This treatment provides an intuitive view of the bias estimation problem. While gyro and accelerometer biases tend to increase the attitude errors, the applied "torques" try to rebalance them. Thus, in the steady-state operation, these "torques" are proportional to biases and can be naturally used to estimate them.

The only tunable parameters of the estimator are three time constants: one for the attitude correction, one for the gyro bias filter, and one for the accelerometer bias filter. Each of them has an obvious meaning and allows the AHRS designer to easily determine the dynamic response properties and to obtain well-predictable estimator behavior.

In the proposed approach, sensor bias observability conditions [4] remain the same as when a Kalman filter is used, as these conditions are completely defined by the system dynamics and available measurements, not by any chosen estimation technique. Nevertheless, these observability conditions are expressed explicitly in terms of the vehicle angular rate, which helps the designer understand and overcome many convergence problems.

## 2. BIAS ESTIMATOR THEORY

Consider a strapdown AHRS that computes the direction cosine matrix $\mathbf{C}$ between the body frame $b = (x_b, y_b, z_b)$ and the "virtual platform" frame $p = (x_p, y_p, z_p)$. The matrix dynamics is described by the equation [4]

$$\dot{\mathbf{C}} = \mathbf{C}\breve{\boldsymbol{\omega}}_b - \breve{\boldsymbol{\omega}}_p \mathbf{C}. \tag{1}$$

Here, $\boldsymbol{\omega}_b$ is the body frame angular rate measured by AHRS gyros; $\boldsymbol{\omega}_p$ is the "virtual platform" angular rate formed by the applied "torques"; an arc denotes a skew-symmetric matrix corresponding to a vector, so that $\breve{\mathbf{a}}\mathbf{b} = \mathbf{a} \times \mathbf{b}$.

Ideally, the "virtual platform" lies in the local level plane, and its axes coincide with the North-East-Down frame $n = (N, E, D)$. Accelerometer measurements $\mathbf{f}_b$ are projected onto the "platform":

$$\mathbf{f}_p = \mathbf{C}\mathbf{f}_b, \tag{2}$$

so that the horizontal components of $\mathbf{f}_p = [f_N, f_E, f_D]^T$ contain no projections of gravity vector $\mathbf{g}$. If this condition is violated, a discrepancy appears between the measured specific force $\mathbf{f}_p$ and the true specific force $\mathbf{f}_n^{ext}$ provided by an external aid. Then the "platform" is "torqued" with the angular rate proportional to that discrepancy,

$$\boldsymbol{\omega}_p = -\breve{\mathbf{k}}_p (\mathbf{f}_n^{ext} - \mathbf{f}_p) \tag{3}$$

where $\mathbf{k} = -k\mathbf{g}/g$ and $k$ is the attitude correction gain.

For a perfect AHRS installed on a vehicle that moves on a flat non-rotating Earth, $\boldsymbol{\omega}_p = \mathbf{0}$; any violations of this condition are due to sensor errors only. This simplification, though not generally applicable to high-performance inertial navigation systems, is valid for low-cost MEMS-based AHRSs where gyro biases are significantly greater than the Earth's angular rate.

In the presence of gyro and accelerometer biases, $\mathbf{b}_g$ and $\mathbf{b}_a$, the "platform" tilt error $\boldsymbol{\theta}$ appears, so the direction cosine matrix variation can be expressed as $\delta\mathbf{C} = \breve{\boldsymbol{\theta}}_p \mathbf{C}$. By varying (1) and substituting $\delta\boldsymbol{\omega}_b = \mathbf{b}_g$, one can get

$$\dot{\boldsymbol{\theta}}_p = \mathbf{C}\mathbf{b}_g - \delta\boldsymbol{\omega}_p. \tag{4}$$

Equation (4) has a clear physical meaning. The rate of change of the attitude error vector is determined by two opposite factors: the gyro biases (projected onto the "virtual platform") and the correction "torques". However, this equation is not convenient for the analysis of strapdown AHRS performance, since its terms are not constant during the vehicle turns. Instead, they are modulated by the $\mathbf{C}$ matrix, even if the sensor biases, $\mathbf{b}_g$ and $\mathbf{b}_a$, are constant. To avoid this problem, (4) can be transformed to the body frame:

$$\dot{\boldsymbol{\theta}}_b + \breve{\boldsymbol{\omega}}_b \boldsymbol{\theta}_b = \mathbf{b}_g - \mathbf{C}^T \delta\boldsymbol{\omega}_p = \mathbf{b}_g - \mathbf{u}. \tag{5}$$

It is the last term $\mathbf{u} = \mathbf{C}^T \delta\boldsymbol{\omega}_p$ that will be used for the estimation of inertial sensor biases. The relation between $\mathbf{u}$ and $\mathbf{b}_g$ is already given by (5). To establish a similar relation between $\mathbf{u}$ and $\mathbf{b}_a$, one should vary (2) with $\delta\mathbf{f}_b = \mathbf{b}_a$:

$$\delta\mathbf{f}_p = \breve{\boldsymbol{\theta}}_p \mathbf{f}_p + \mathbf{C}\mathbf{b}_a \tag{6}$$

and then substitute (6) into (3):

$$\delta\boldsymbol{\omega}_p = \breve{\mathbf{k}}_p \breve{\boldsymbol{\theta}}_p \mathbf{f}_p + \breve{\mathbf{k}}_p \mathbf{C}\mathbf{b}_a =$$
$$= \left(\mathbf{k}_p^T \mathbf{f}_p\right)\boldsymbol{\theta}_p - \left(\mathbf{k}_p^T \boldsymbol{\theta}_p\right)\mathbf{f}_p + \breve{\mathbf{k}}_p \mathbf{C}\mathbf{b}_a. \tag{7}$$

The first term in the right-hand side of (7) is proportional to the attitude error to be eliminated. It is the core of the whole correction scheme, which provides the desired error feedback. The second term is an additional error that appears when the vectors $\mathbf{k}$ and $\boldsymbol{\theta}$ are not orthogonal. When only tilt errors exist, $\boldsymbol{\theta}$ lies in the level plane, while $\mathbf{k}$ is always vertical by its definition. Thus, the term is nonzero only when a heading error is present. The third term reflects the influence of accelerometer biases on the correction accuracy. Though generally harmful, this term allows for the estimation of accelerometer biases by means of the $\mathbf{u}$ signal.

Finally, (7) can be transformed to the body frame and substituted into (5) to get the general error equation

$$\dot{\boldsymbol{\theta}}_b = -\breve{\boldsymbol{\omega}}_b\boldsymbol{\theta}_b + \mathbf{b}_g - \left(\mathbf{k}_p^T\mathbf{f}_p\right)\boldsymbol{\theta}_b + \left(\mathbf{k}_p^T\boldsymbol{\theta}_p\right)\mathbf{f}_b - \breve{\mathbf{k}}_b\mathbf{b}_a. \quad (8)$$

For a GPS-aided AHRS installed on a land vehicle, some simplifications in (8) can be made. First, vertical accelerations are typically small, so $f_D = -g$ and $\mathbf{k}_p^T\mathbf{f}_p = kg$. Second, the true heading is always defined by the vehicle velocity direction, so the heading error can be held zero as long as the vehicle is in motion and navigation satellites are visible. Hence, $\mathbf{k}_p^T\boldsymbol{\theta}_p = 0$. Third, roll and pitch angles never exceed several degrees, so $\mathbf{k}_b = \mathbf{k}_p$. Under these assumptions, one can rewrite (8) in the form

$$\dot{\boldsymbol{\theta}}_b + (\breve{\boldsymbol{\omega}}_b + kg\mathbf{I})\boldsymbol{\theta}_b = \mathbf{b}_g - \breve{\mathbf{k}}_p\mathbf{b}_a. \quad (9)$$

Equation (9) shows that the attitude correction loop can be described as a low-pass filter excited by sensor biases. This result is well-known for gimbaled AHRSs [13], except for the appearance of $\breve{\boldsymbol{\omega}}_b$ matrix in the left-hand side of (9), which is specific for the strapdown mechanization. This matrix plays an important role in the separation of gyro and accelerometer biases in the estimation procedure.

Equation (9) has a remarkable property: its right-hand side is almost constant irrespective of any changes in vehicle heading. Therefore, the correction gain $k$ can be chosen in such a way that the filter (9) converges to a steady state where $\dot{\boldsymbol{\theta}}_b = \mathbf{0}$ and

$$\boldsymbol{\theta}_b = (\breve{\boldsymbol{\omega}}_b + kg\mathbf{I})^{-1}(\mathbf{b}_g - \breve{\mathbf{k}}_p\mathbf{b}_a). \quad (10)$$

When the solution is substituted into (5), a steady-state "torque" can be expressed as

$$\mathbf{u} = \mathbf{b}_g - \breve{\boldsymbol{\omega}}_b(\breve{\boldsymbol{\omega}}_b + kg\mathbf{I})^{-1}(\mathbf{b}_g - \breve{\mathbf{k}}_p\mathbf{b}_a). \quad (11)$$

Equation (11) gives a linear relationship between the "torque", treated as the measurement, and sensor biases treated as estimated states. Since there are two states in a single measurement equation, some additional observability considerations are needed. First, suppose that a vehicle is moving along a straight line. In this case, $\boldsymbol{\omega}_b = \mathbf{0}$ and

$$\mathbf{b}_g = \mathbf{u}. \quad (12)$$

Thus the applied "torque" merely compensates the gyro bias and can be used to estimate it. Accelerometer biases are not observable in these conditions.

Second, suppose that the gyro bias has already been estimated and compensated, so that $\mathbf{b}_g = \mathbf{0}$, and the vehicle is turning at the angular rate $\boldsymbol{\omega}_b$. Equation (11) then takes the form

$$\breve{\boldsymbol{\omega}}_b(\breve{\boldsymbol{\omega}}_b + kg\mathbf{I})^{-1}\breve{\mathbf{k}}_p\mathbf{b}_a = \mathbf{u}, \quad (13)$$

which can be further solved for the unknown accelerometer biases. The solution has an extremely simple expression if only the vertical component of the vehicle's angular rate, $\omega_{zb}$, is considered:

$$\mathbf{b}_a = g\begin{bmatrix} 1/\omega_{zb} & -\tau & 0 \\ \tau & 1/\omega_{zb} & 0 \\ 0 & 0 & 0 \end{bmatrix}\mathbf{u} \quad (14)$$

where $\tau = 1/(kg)$ is the attitude correction time constant.

Equations (12) and (14) provide bias estimates for the roll and pitch inertial sensors in terms of the residual "torques" applied to the "virtual platform" of the AHRS. They constitute the core of the proposed estimation approach (Fig. 1).

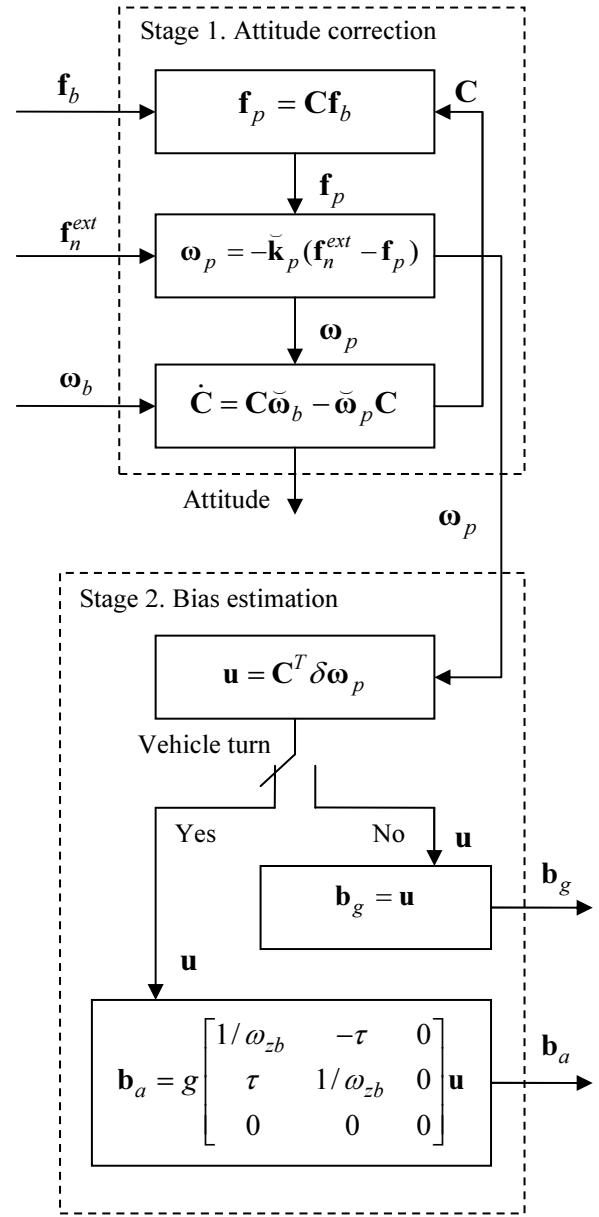

Fig. 1. Bias estimator block diagram.

The main advantage of this approach over traditional Kalman filter-based methods is its simplicity: while the complete system model contains nonlinear dynamics (1) – (3), the estimation equations (12) and (14) are given in a *closed form*. Moreover, they do not require any high-dimensional vector computations.

Even though the estimation equations can be used directly, it is likely that the "torque" **u** is very noisy. That is why it is recommended that two simple first-order low-pass filters be implemented: one for the gyro biases, and another for the accelerometer biases.

The obtained estimates can be used in two different ways. The first way is to augment the attitude estimation part by velocity and position correction blocks and to construct bias estimate feedback loops. In this case, the estimator becomes a self-contained navigation filter with the same capabilities as the Kalman filter. The second way is to use the bias estimator as an independent "black box" whose estimates are subtracted from the input data of a primary navigation filter.

## 3. BIAS ESTIMATOR IMPLEMENTATION AND TESTING

The proposed estimator was implemented by Topcon Positioning Systems, LLC, in the software of a GPS/GLONASS receiver equipped with a built-in low-cost MEMS inertial measurement unit (IMU). The aim was to estimate and compensate roll and pitch gyro biases of the order of 0.1 deg/s and accelerometer biases of the order of 0.2 m/s$^2$.

To obtain true acceleration data, the carrier-phase velocity measurements provided by the receiver were differentiated. The attitude correction time constant was set to 4 s, whereas time constants for both bias filters were set to 40 s.

The tests were conducted on a John Deere 5515 wheel tractor and a Caterpillar Challenger rubber tracked tractor with the receiver units installed on the cabin roofs. Test paths consisted of straight line segments as well as of turn arcs to satisfy the observability conditions for all estimated quantities (Fig. 2).

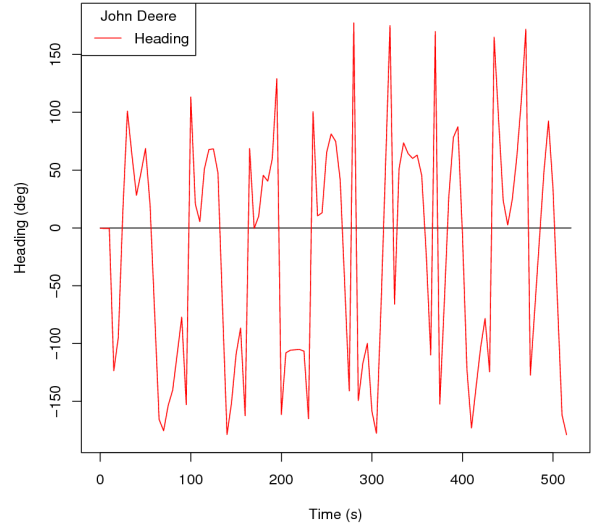

Fig. 2a. Heading angle (John Deere wheel tractor).

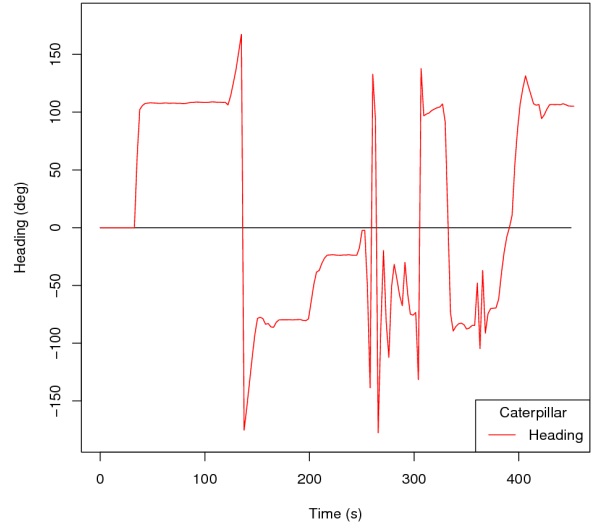

Fig. 2b. Heading angle (Caterpillar tracked tractor).

Gyro bias estimates obtained in the tests are shown in Fig. 3. To assess the estimation accuracy, these estimates were compared to the true bias values obtained at rest immediately after the test by direct averaging of gyro measurements.

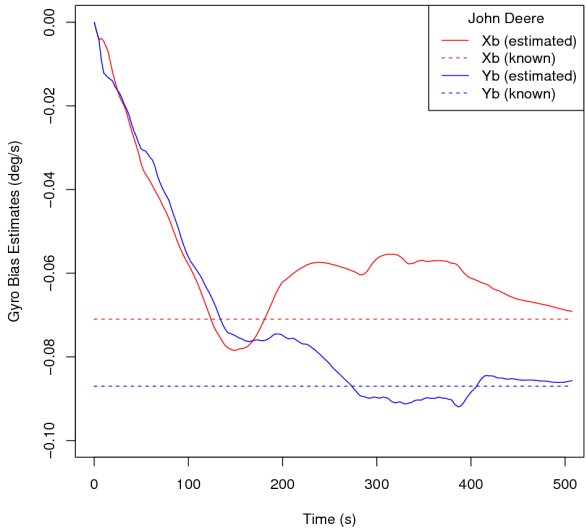

Fig. 3a. Gyro bias estimates (John Deere wheel tractor).

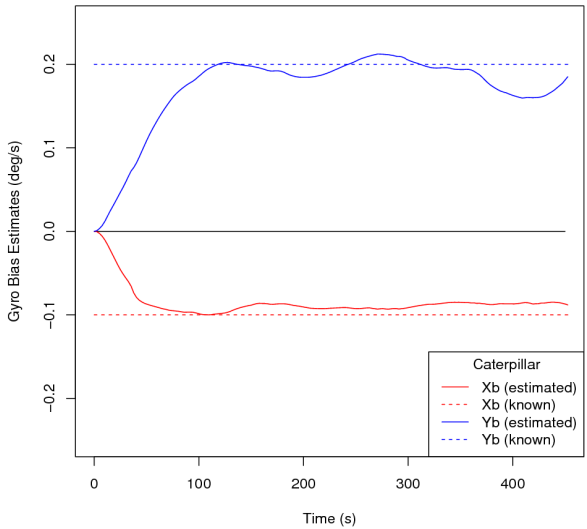

Fig. 3b. Gyro bias estimates (Caterpillar tracked tractor).

The gyro bias filter convergence time was about 120 s, which is equal to three time constants, as predicted by filtering theory. The steady-state estimation errors reached 0.01 deg/s (RMS) in the first test and 0.02 deg/s (RMS) in the second one. These results can be compared to the performance of a "symmetry-preserving filter" where bias estimate instabilities were of the order of 0.01 rad/s, or 0.5 deg/s [14].

Accelerometer bias estimation is usually considered as a much more difficult problem because these biases are not observable in the straight motion of a vehicle [4]. Bias estimates obtained in both tests were less than 0.04 m/s$^2$. The true bias values were unknown, as they were undistinguishable from the tractor's actual attitude and IMU installation errors. Therefore the estimation accuracy assessment was done indirectly. The precisely known artificial biases were added to the accelerometer measurements, and then the tests were repeated. In Fig. 4, the obtained estimates and the artificial biases are shown.

The accelerometer bias filter convergence time was about 300 s, which was more than twice longer than for the gyro bias filter in spite of the fact that both filters had equal time constants. This reflected the weak observability of accelerometer biases. The steady-state estimation errors were 0.04 m/s$^2$ (RMS) in both tests.

The tests also confirm that the estimation technique is not sensitive to IMU installation angles if they do not exceed 10…15 deg. Any installation error is implicitly treated as a redefinition of body frame axes and does not affect the estimation accuracy.

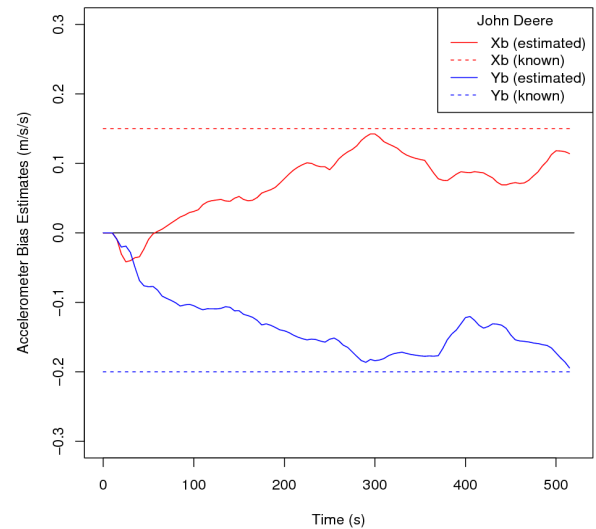

Fig. 4a. Accelerometer bias estimates (John Deere wheel tractor).

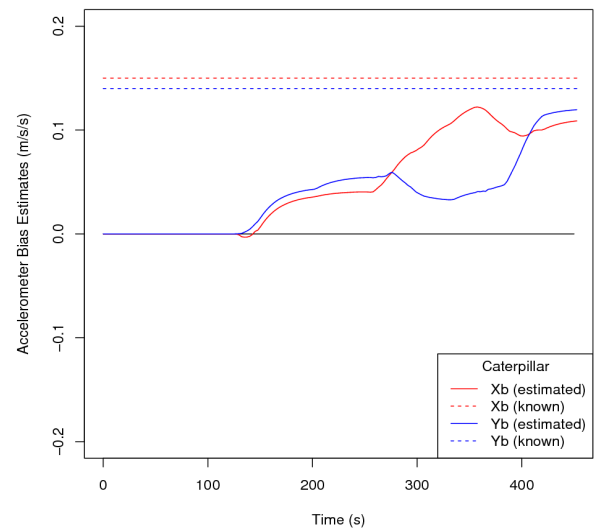

Fig. 4b. Accelerometer bias estimates (Caterpillar tracked tractor).

## 4. CONCLUSION

The proposed sensor bias estimator proved its feasibility for the estimation of MEMS gyro and accelerometer biases of a low-cost AHRS installed on a land vehicle. It can be widely used in the fields of precision agriculture, construction engineering, etc. The applicability of this approach to tactical grade inertial systems requires further investigation. In this case, the attitude correction gain should be significantly decreased, and the Earth's shape and rotation should be appropriately taken into account. For navigation grade inertial systems the more general techniques, such as Kalman filtering, are probably still preferable.